\documentclass[10pt,conference]{IEEEtran}
\IEEEoverridecommandlockouts
\usepackage{cite}
\usepackage{amsmath,amssymb,amsfonts}
\usepackage{algorithmic}
\usepackage{graphicx}
\usepackage{textcomp}
\usepackage[table,xcdraw]{xcolor}
\usepackage[inline]{enumitem}
\usepackage{url}

\usepackage[T1]{fontenc}
\usepackage[utf8]{inputenc}
\usepackage{nicefrac}

\usepackage{siunitx}  
\usepackage{flushend}

\usepackage{booktabs}
\SetEnumitemKey{midsep}{topsep=3pt, partopsep=0pt}
\newif\iftodos
\todostrue
\todosfalse

\iftodos
\newcommand{\PP}[1]{\todo[inline]{#1}}
\newcommand{\JN}[1]{\todo[inline,color=green!40]{JN: #1}}
\newcommand{\jn}[1]{\todo[color=green!40,size=\tiny]{JN: #1}}
\else
\newcommand{\PP}[1]{}
\newcommand{\JN}[1]{}
\newcommand{\jn}[1]{}
\fi

\def\BibTeX{{\rm B\kern-.05em{\sc i\kern-.025em b}\kern-.08em
    T\kern-.1667em\lower.7ex\hbox{E}\kern-.125emX}}
\begin{document}

\title{The Secret Life of CVEs}


 \author{\IEEEauthorblockN{1\textsuperscript{st}Piotr Przymus}
 \IEEEauthorblockA{\textit{Nicolaus Copernicus University} \\
 Toru\'{n}, Poland\\
 {\small ORCID 0000-0001-9548-2388}\\
piotr.przymus@mat.umk.pl}
 \and
 \IEEEauthorblockN{2\textsuperscript{nd} Miko\l{}aj Fejzer}
 \IEEEauthorblockA{\textit{Nicolaus Copernicus University}\\
 Toru\'{n}, Poland \\
 {\small ORCID 0000-0003-1496-2289}\\
mfejzer@mat.umk.pl}
 \and
 \IEEEauthorblockN{3\textsuperscript{rd} Jakub Nar\k{e}bski}
 \IEEEauthorblockA{\textit{Nicolaus Copernicus University} \\
 Toru\'{n}, Poland\\
 {\small ORCID 0000-0002-3296-3915}\\
jakub.narebski@mat.umk.pl
}
 \and
 \IEEEauthorblockN{4\textsuperscript{th} Krzysztof Stencel}
 \IEEEauthorblockA{\textit{University of Warsaw}\\
 Warsaw, Poland \\
 {\small ORCID 0000-0001-6356-4872}\\
stencel@mimuw.edu.pl
}
 }

\maketitle

\begin{abstract}
The Common Vulnerabilities and Exposures (CVEs) system is a reference method for documenting publicly known information security weaknesses and exposures.
This paper presents a study of the lifetime of CVEs in software projects and the risk factors affecting their existence.
The study uses survival analysis to examine how features of programming languages, projects, and CVEs themselves impact the lifetime of CVEs.
We suggest avenues for future research to investigate the effect of various factors on the resolution of vulnerabilities.
\end{abstract}

\begin{IEEEkeywords}
CVE, Mining software repositories, Software quality, Survival analysis
\end{IEEEkeywords}


\section{Introduction}
\label{sec:intro}

Vulnerabilities in software systems can be exploited by attackers to gain unauthorized access or cause other harm.
Common Vulnerabilities and Exposures (CVEs) \cite{cve-db} are publicly disclosed vulnerabilities that have been assigned unique identifiers in the CVE system.
Understanding the lifetime of a CVE in a software project helps organizations and developers to better identify and address potential security weaknesses.

Survival analysis \cite{liu2012survival} is a statistical method that can be used to investigate the time until an event of interest occurs.
In the context of software security, the resolution of a vulnerability can be considered as the event of interest, while the duration of the period until resolution is the time variable. 
In consequence, we can estimate the probability that a vulnerability has not been resolved at a given time, providing insight into the lifetime of a CVE in a software project.

In this paper, we will investigate the lifetime of CVEs in software projects, as well as the risk factors that may influence it. 
These risk factors include features related to the characteristics of programming languages, the CVE itself, and the project.
We will study the following research questions ({\bf RQ}):
{\bf RQ1:} What are the characteristics of the CVE lifetime within a project?
{\bf RQ2:} What are the risk factors affecting the lifetime of a CVE in a project, considering the characteristics of the used programming languages, the unique features of the CVE, and the characteristics of the project?
The final dataset, along with the source code and notebooks used to extract and analyze the data, are accessible on Figshare~\cite{dataset}.

\textbf{Preliminaries:} Survival analysis \cite{liu2012survival} is particularly useful for studying time-to-event data, where the event of interest is e.g., the failure of a mechanical component or the resolution of a vulnerability.
Assume a random variable represents the time of the studied event.
The \emph{survival function} $S(t)$ is the probability that up to time $t$ the event of interest did not occur.
If $F$ is the cumulative distribution function of this random variable, then the survival function is\ $S(t) = 1 - F(t)$.

In this article we estimate the survival function using the Kaplan-Meier estimator \cite{doi:10.1080/01621459.1958.10501452}.
If $d_i$ is the number of events that happened at $t_i$, and $n_i$ is the number of individuals that survived up to $t_i$, than Kaplan-Meier estimate the survival function as:
$\hat{S}(t) = \prod_{i:t_i \le t} (1 - \nicefrac{d_i}{n_i})$.

In order to relate obtained random variables we use Sommers' D measure, also known as $D_{xy}$ \cite{newson2010comparing}.
It evaluates the ordinal association between two possibly dependent variables. 
Sommers' D assumes values between $-1$ and $+1$.
The former value indicates that all pair of the variables agree, and the latter implies complete disagreement. We typically look for high values of $|D_{xy}|$, as this indicates improved predictive ability of examined factor.

\section{Methods}
\label{sec:methods}
{\bf Initial dataset and pre-processing.} 
In this study, we merged two datasets, the World of Code (WoC)~\cite{DBLP:journals/ese/MaDBAVTKZM21, msr-challenge} and the Common Vulnerabilities and Exposures (CVE) database~\cite{nvd, cve-db}. The WoC dataset version U contains the data from 173 million git repositories and over 3.1 billion commits, retrieved on 28-11-2021, while the CVE database, retrieved on 26-07-2022, was obtained using the cve-search project \cite{Cvesearch2023} that integrates various sources, including~\cite{nvd, cve-db}. 
The data sources and actions taken are shown in the~Fig. 1.h.

From the WoC dataset we extracted \num{702786} commits related to CVE fixes by matching CVE-ID in the commit messages.
We used the WoC "c2P" map to retrieve the projects for each commit (omitting project forks).
For each project we retrieved corresponding metadata from WoC database.
Next we combined the commit and project data with CVE details.

We extracted the file extensions of modified files in each commit in order to identify the programming languages used. Using the GitHub Linguist library~\cite{github-linguist}, we limited the file extensions by matching the programming-related extensions. We calculated the total number of modified files in popular programming languages (based on the TIOBE top 100 ranking~\cite{index2022tiobe}) for each project and CVE pair. All other programming related extensions were grouped as "misc" files.

To ensure the accuracy of our analysis, we filtered commits based on project and commit date criteria.
In cases where multiple forks were present but no root project was identified, we used the fork with the biggest number of commits as the root project.
Any commits without a recognizable project were removed from the dataset.
We further filtered commits with creation dates before 1999 or after 2022, resulting in \num{643319} commits for analysis, accounting for $91.4\%$ of the total found CVE-related commits. With these filtered commits, we calculated the lifetime of each CVE.

{\bf CVE lifetime definition.} 
The concept of the lifetime of CVE in a given project refers to its presence from its introduction to its resolution limited to given project. 
This can include the following stages:

\begin{description}[leftmargin=0.2cm]
\item[1. Introduction] -- a vulnerability is introduced into the project, by a coding error or other means.
\item[2. Discovery] -- a vulnerability is discovered, the process of addressing it can begin.
\item[3. CVE ID] -- a unique CVE ID identifier is assigned to it.
\item[3* \hspace{-2pt}Embargo] -- optional period before the details of the CVE are publicly disclosed, until a fixing patch is released.
\item[4. CVE Publication] -- publication of CVE details, including information on affected software, severity, and description.
\item[5. Resolution] -- a set of commits addressing the CVE in the project, occurring anytime after discovery.
\end{description}

The phases of the CVE lifetime and its relationship with commits, projects, and embargos is depicted in the CVE Lifetime section of the accompanying infographic (see Fig.~1.f).
We conduct a large-scale study that focuses solely on projects and commits that use CVE IDs during the development process.
Developers may include CVE ID in commit message for tracking purposes, for reasons like fixing a vulnerability or tracking details for data mining, software screening, penetration testing, or proof-of-concept.
We do not differentiate between these cases as there is no reliable way to do so without manual inspection.

For the purpose of this analysis we use two observable points: $\mathbf{T}_s$ - Start of the observation, global value computed as $\min(C_{glob}, C_{pub})$ where $C_{glob}$ is the minimal date of all commits mentioning given CVE ID, and $C_{pub}$ is the publication date. 
$\mathbf{T}_e$ - end of the observation, that is maximal date of all commits mentioning given CVE for given project. 
Then CVE Liftime is defined as $Y=\mathbf{T}_e-\mathbf{T}_s$ and is computed in days.
By definition, our observations are right-censored, meaning that the lifetime of a CVE in a project may exceed the computed value, but the exact amount is unknown as we do not observe the Introduction or Discovery.
Our definition of CVE Lifetime matches the period where possibly unfixed bug is publicly visible as security vulnerability.

To enhance relevancy and minimize information noise, we confined the survival analysis to project-CVE pairs where bug fixes occurred within 365 days of the CVE publication date.

\section{Results}
\label{sec:results}
\noindent{\bf RQ1:} What are the characteristics of the CVE lifetime within a project?

To address this issue, we grouped the set of commits by CVE-ID and project, then computed the lifetime of the CVE in the project and calculated basic statistics about the embargo, the interaction between programming languages, etc. (see the {\bf RQ1} in infographic, Fig.~1.a-d and Fig.~1.i).
 
The analyzed cohort of 62.8K objects, which includes unique 9.8K projects and 22.7K CVEs, is a subset of the total number of selected commits associated with over 16K projects and 27K CVEs, resulting in 97.7K watched objects (see Fig.~1.a). The subset was selected based on time time horizon for CVE lifetime and $Y\leq 365$ was adopted.

The median number of commits needed to fix a CVE is 1, with the 75th percentile being 3. While most CVE fixes require a small number of commits, some projects differ significantly, such as operating system source projects that aggregate software from other projects. The median time to fix a CVE is 34 days, with a $20\%$ chance of it not being addressed after 150 days.
Embargoed CVE repairs were observed, with 22.8K made during the embargo and 16.5K after it. The study considered CVE embargoed if any related commit was seen before publication date of CVE. For 23.5K, no embargo was found, but it cannot be ruled out.

An intriguing finding is the large number of projects that use multiple programming languages in the repair process (see Fig.~1.g). Most CVE fixes involve changes to code written in more than one programming language, or to supporting ("misc") files for the project. The lowest percentage of mixed languages fixes are for C and Java ($\approx57\%$). Other languages have a higher proportion of mixed language fixes ($\geq86\%$).
We analyzed language coexistence using a chord chart to understand relationships during a CVE's lifetime.

\noindent{\bf RQ2:} What are the risk factors affecting the lifetime of a CVE in a project, considering the characteristics of the used programming languages, the unique features of the CVE, and the characteristics of the project?

To address this question, we utilized the $\hat{S}(t)$ survival function estimate and its plots. 
We compared impact of individual features using Sommers' D measure ($D_{xy}$), with confidence intervals (CI) determined via bootstrap analysis ($n\!=\!100$), results within $95\%$ CI and spread $\leq\pm 0.001$.
The in-depth results of the analysis can be found in the infographic, table and charts designated as {\bf RQ2} (see Fig.~1.e and Fig.~1.j-m)

\begin{figure*}[p]
  \centering
  \includegraphics[width=0.915\textwidth]{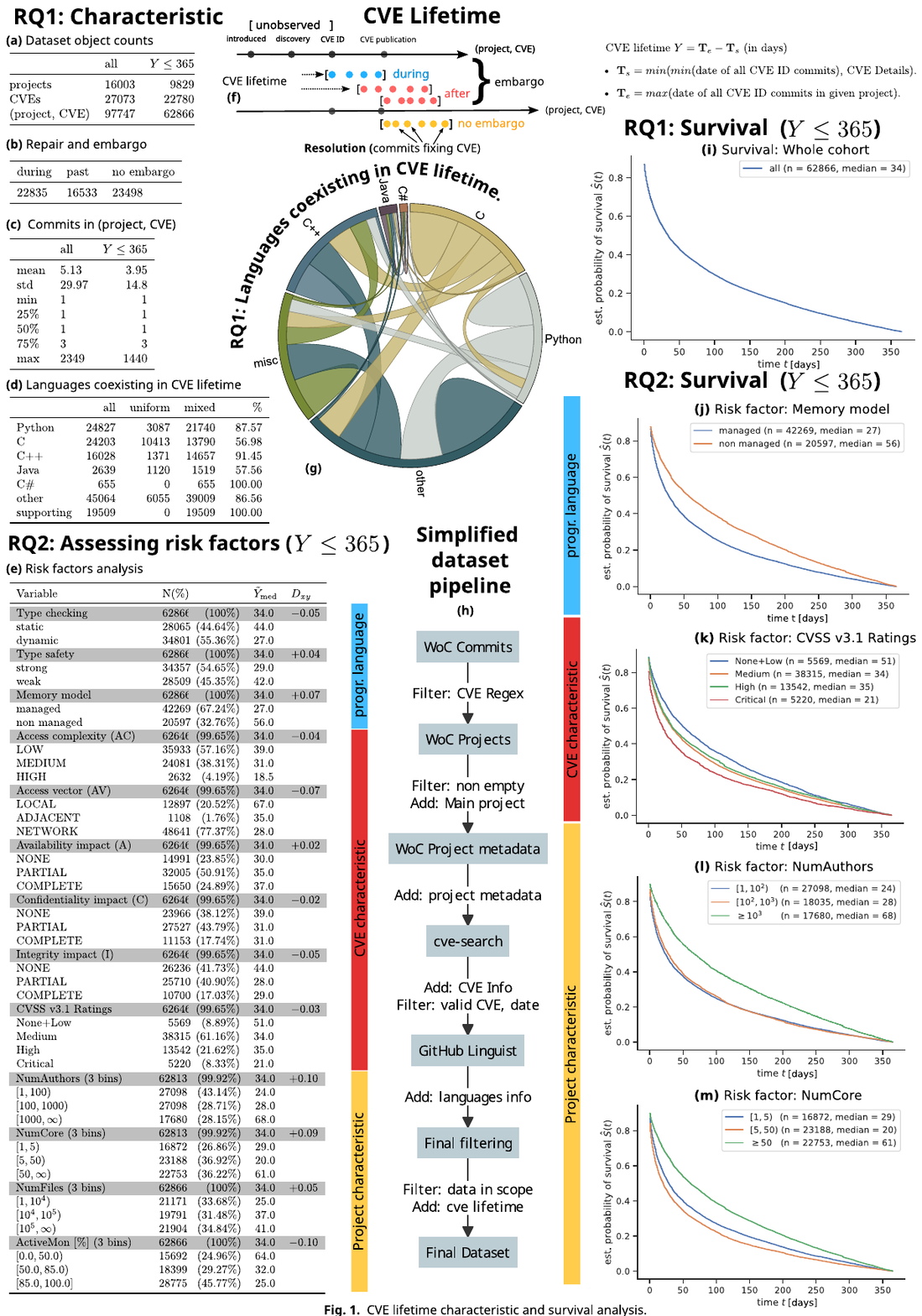}
\end{figure*}

\noindent {\bf Programming language characteristics.}
To minimise effects of multiple languages present in the lifespan of a CVE on each other, we have decided to designate a primary language for each CVE-project pair, determined by the language for which most files were changed.
Subsequently, we assigned the features to all programming languages in the \cite{index2022tiobe} ranking, using the list of features from \cite{wikipedia-types} as a data source.
We selected basic features: type checking, type safety, and memory model, as these could easily be attributed to the respective languages (see Fig.~1.e).
The results show that the memory model has the highest importance ($D_{xy}\!=\!0.07$) with a median of 27 days for managed and 56 days for non-managed (see Fig.~1.e and Fig.~1.i). The results for type checking were little surprising, with importance ($D_{xy}\!=\!-0.05$) and a median of 27 days for dynamic and 44 days for static.
Type safety had the lowest importance ($D_{xy}\!=\!0.04$) with a median of 29 days for strong and 42 days for weak.
Results mostly align with expectations.

\noindent\textbf{CVE characteristics.}
The lifetime of a CVE within a project can be impacted by its characteristics, like features described in~\cite{cvss2.0}. We use the qualitative metrics (see Fig.~1.e), such as access complexity (AC), access vector (AV), availability impact (A), confidentiality impact (C), and integrity impact (I).
AV has the greatest impact, with a $D_{xy}\!=\!0.07$, where attacks from the network have a median resolution time greater than 28 days and local attacks have a median of 67 days. The next factor is I, with $D_{xy}=-0.05$, and a median resolution time of 44 days for NONE, 28 and 29 days for PARTIAL and COMPLETE, respectively. 
AC ($D_{xy}=-0.04$) is surprising, with LOW having 39 days median time, MEDIUM 31 days, and HIGH 18 days. Further analysis may be needed as LOW indicates elevated risk in this case. The impact of other factors on survival time is minimal, per the risk table.

For CVSS Score v3.1\cite{cvss3.1}, the impact is $D_{xy} = -0.03$, with bugs marked as critical being resolved faster, next HIGH and MEDIUM bugs are fixed, with LOW bugs being fixed slowest as expected (see Fig.~1.k). The indistinctiveness between HIGH and MEDIUM bugs calls for a deeper analysis.

\noindent{\bf Project characteristics.}
The greatest impact on CVE lifetime can be seen from project characteristics, including the ratio of months with at least one commit compared to project lifetime (ActiveMon [\%]), the number of core developers (NumCore), the number of all unique authors (NumAuthors), and the total number of files in the project (NumFiles).
Bug fix time increases with number of authors in project (NumAuthors, $D_{xy}=0.10$), median of 24-28 days for projects with less than 1K authors, 68 days for more authors involved (see Fig.~1.l).
Observing only core developers (NumCore, $D_{xy}\!=\!0.09$), medium projects (5-50 developers) have fastest median time of 20 days, then small teams ($\leq 5$ developers, median 29 days), then large teams with 61 days to fix (see Fig.~1.m).
Another good indicator, ActiveMon [\%] ($D_{xy}\!=\!-0.10$), shows that active projects fix issues quicker (median 25 days) than less active ones (median 64 days).
The impact of the NumFiles ($D_{xy}\!=\!0.05$) is less significant, it can be seen that the bug fix time increases with the increase in the number of files, with a median of 25 days for projects with $\leq 10K$ files, and 41 days for projects with more than 100K files.

\section{Threats to validity}
\label{sec:validity}

As this is a large case study there are natural limitations.
\begin{enumerate*}
\item {\it Observability of CVE in project} is one of the biggest limitations of this research.
We rely on developer meticulousness to put CVE ID in commit messages.
Thus we have no means to reliably spot {\it (partial) silent repairs} --- situation were developer fixes CVE withholding CVE ID information in (some) commit messages.
Additionally we cannot detect CVE if commit message was created via squash merge without including all details.
This limitation results in {\it selection bias} for {\it non silent repairs}.

\item {\it Data quality} depends on the correctness of the data sources we rely on, thus anomalies in data may influence this analysis.
We countered this by filtering suspicious data, like commit ``time travellers'' (commits with suspicious date from past or future with respect to range of dates that are covered by WoC and CVE database, as well as commits that fixes CVEs more than one year before CVE publication date).
Additionally we removed each commit which corresponding project or CVE metadata were not available.
Thus we excluded \num{59467} (8.6\%) of CVE related commits due to wrong or missing data.

\item {\it Right-censored data}. We assume that all our observations are right censored as we have no means to mass check CVE introduction and discovery dates.
Moreover, due to exclusion of commits without explicit CVE ID our observed start of CVE lifetime can be inaccurate, which introduces {\it measurement error}.

\item {\it Confounding variables}. There may be other variables not considered in the analysis that are influencing the lifetime of the CVE.
  
\item {\it Embargo estimation}. Due to the observability the definition of an embargo adopted in this work may be inadequate.

\item {\it Ethical implications}.
    We do not process any sensitive individual contributor data, nor make automated judgments to ascribe characteristics to individuals.
\end{enumerate*}

\section{Related work}
\label{sec:related}

Survival analysis was applied to lifecycles of vulnerabilities and bugs by numerous researchers in empirical studies discussed below.

\begin{enumerate*}
\item 
Wedel et al. \cite{DBLP:conf/esem/WedelJG08} employed survival analysis to analyse the Eclipse project. 
They found that the code size was the most significant factor for defect duration in a release.

\item 
In a survival analysis of bugs across 4 projects~\cite{DBLP:conf/wcre/CanforaCCP11}, 
the authors linked the bug's lifespan to specific code syntax and identified those related to long-lived bugs.

\item 
A survival analysis of issues was carried out in a study of selected GitHub projects \cite{DBLP:conf/socinfo/JarczykGJBW14}. 
Its authors analysed \num{2000} projects for issue survival.
They used binomial regression agaist approximately a dozen basic project property features.

\item 
Survival of vulnerabilities in Android was a subject of the study \cite{DBLP:conf/msr/VasquezBE17}. 
This study limited to this operating system indicates that more severe vulnerabilities are those that live longer in contrast to our findings.
The authors use a similar schema of vulnerability life.
However, they try to identify when the vulnerability was \emph{introduced}, 
while in our rather black-box approach we consider only the \emph{reporting} time.

\item 
Tools of survival analysis similar to our toolkit were used in the article \cite{farris2017vulnerability}.
However the authors took another point of view, analysing installed and running software on servers, via monthly scan of \num{2000} machines using Nessus software for 12 months.
They do not trace vulnerabilities in the sources as our study does.

\item 
Householder et al. \cite{doi:10.1080/01621459.1958.10501452} analysed public exploit availability
in ExploitDB and Metasploit framework between 2013 and 2020 on \num{75807}
software vulnerabilities with assigned CVE-ID.
The authors used survival analysis via Kaplan-Meier estimator, using exploit publish date as patient death, to discover how CVE-ID features influence exploit preparation.
While our approach focuses on vulnerable source code of an application, with median time of 34 days to create patch,
Householder et al. focuses on tools available on the attacker side, showing that within 28 days 75\% CVE has available exploit.

\item 
Nappa et al.~\cite{DBLP:conf/sp/NappaJBCD15} researched deployment of software patches to hosts.
They used survival analysis on combined patch delay and patch deployment defined as the time between vulnerability disclosure and time of patch arrival to user hosts (software update).
The authors used 10 popular client-side applications affected by \num{1593} vulnerabilities.
They used data from 8.4 million hosts over a span of 5 years.
Their method differs from our approach, as we do not track installed program versions nor user reports (patch deployment), but we focus on the patching process in the source code (patch delay).
According to \cite{DBLP:conf/sp/NappaJBCD15} 92\% vulnerabilities are patched within 30 days around disclosure.
On the other hand when we analyzed 
WoC dataset~\cite{DBLP:journals/ese/MaDBAVTKZM21, msr-challenge} enriched with CVE data~\cite{nvd, cve-db},
we observed that the median time for CVE fix is 34 days.

\item 
Bilge et al.~\cite{DBLP:conf/ccs/BilgeD12} analyzed data from 11 million hosts between 2008 and 2011 via Symantec Worldwide Intelligence Network Environment platform to prepare a method of identifying zero day attacks.
The authors aimed at finding executable binary files on user hosts used to exploit vulnerabilities before their public disclosure.
Their results are not directly comparable to ours, 
as we focus on software patch side, and cannot observe attacker activities.

\item 
Frei et al.~\cite{DBLP:conf/sigcomm/FreiMFP06} conducted analysis of 14,000 vulnerabilities published between 1996 and 2006 
to determine the date of the discovery, disclosure, exploit and patch of each one.
They correlated CVE entries with 80,000 additional sources.
The authors found that 95\% of the exploits are available within a month of disclosure of the vulnerability, while between 55\% to 85\% of the patches are available upon same date or next 30 days.
The patch availability aligns with our result of 34 days of patch creation.

\end{enumerate*}

Our study is more comprehensive due to its utilization of a vast dataset from WoC~\cite{DBLP:journals/ese/MaDBAVTKZM21,msr-challenge} and a comprehensive database of all CVE vulnerabilities.
Furthermore, we reveal a greater quantity of features that influence the survival of bugs/CVEs.


\section{Conclusions}
\label{sec:conclusions}
We found interesting results.
With regards to RQ1, the median time to fix a CVE was 34 days, with 75\% of fixes requiring only 1 to 3 commits.
Most recent CVE fixes involved changes to code written in multiple programming languages or related files.
Regarding RQ2, all factors had a value of $|D_{x,y}|\leq 0.104$. The memory model was found to be the most important characteristic of programming languages, while Access Vector had the greatest influence on CVE characteristics.
The project characteristics had the biggest impact on bug fix time, with more authors resulting in slower, and being more active resulting in quicker response times. The effect of the number of files was less pronounced but consistent with expectations.
Some interesting and unexpected results were observed and warrant further investigation.
Due to space limitations, only a selective overview is presented here, but a comprehensive examination should be considered in the future.


\bibliographystyle{unsrt}
\bibliography{biblio}

\end{document}